\documentclass[journal,draftcls,onecolumn,12pt,twoside]{IEEEtranTCOM}

\normalsize
\usepackage{cite}
\ifCLASSINFOpdf
  \usepackage[pdftex]{graphicx}
  \graphicspath{{../pdf/}{../jpeg/}}
  \DeclareGraphicsExtensions{.pdf,.jpeg,.png}
\else
  \usepackage[dvips]{graphicx}
  \graphicspath{{../eps/}}
  \DeclareGraphicsExtensions{.eps}
\fi
\usepackage[cmex10]{amsmath}
\usepackage{algorithm}
\usepackage{algpseudocode}
\usepackage{array}
\usepackage[caption=false,font=footnotesize]{subfig}

\usepackage{mathrsfs}
\newtheorem{myprop}{Proposition}

\begin{document}

\title{Decoupled Uplink and Downlink in a Wireless System with Buffer-Aided Relaying}

\author{Rongkuan~Liu,~\IEEEmembership{}
        Petar~Popovski,~\IEEEmembership{Fellow,~IEEE,}
        and~Gang~Wang,~\IEEEmembership{Member,~IEEE}%
\thanks{R. Liu is with the Communication Research Center, Harbin Institute of Technology, Harbin 150001, China, and also with the Department of Electronic Systems, Aalborg University, Aalborg 9220, Denmark (e-mail: liurongkuan@hit.edu.cn).}%
\thanks{P. Popovski is with the Department of Electronic Systems, Aalborg University, Aalborg 9220, Denmark (e-mail: petarp@es.aau.dk).}%
\thanks{G. Wang is with the Communication Research Center, Harbin Institute of Technology, Harbin 150001, China (e-mail: gwang51@hit.edu.cn).}}


\maketitle

\begin{abstract}
The paper treats a multiuser relay scenario where multiple user equipments (UEs) have a two-way communication with a common Base Station (BS) in the presence of a buffer-equipped Relay Station (RS). Each of the uplink (UL) and downlink (DL) transmission can take place over a direct or over a relayed path. Traditionally, the UL and the DL path of a given two-way link are \emph{coupled}, that is, either both are direct links or both are relayed links. By removing the restriction for coupling, one opens the design space for a \emph{decoupled} two-way links. Following this, we devise two protocols: orthogonal decoupled UL/DL buffer-aided (ODBA) relaying protocol and non-orthogonal decoupled UL/DL buffer-aided (NODBA) relaying protocol. In NODBA, the receiver can use successive interference cancellation (SIC) to extract the desired signal from a collision between UL and DL signals.  For both protocols, we characterize the transmission decision policies in terms of maximization of the average two-way sum rate of the system. The numerical results show that decoupling association and non-orthogonal radio access lead to significant throughput gains for two-way traffic.
\end{abstract}


\section{Introduction}
\begin{figure}[!t]
\centering
\includegraphics[width=1.6in]{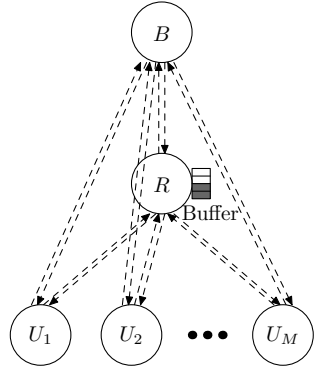}
\caption{Two-way multiuser relay network with buffer.}
\label{cm}
\end{figure}

\IEEEPARstart{T}{he} traffic in broadband wireless networks is essentially two-way, featuring both uplink (UL) and downlink (DL) transmissions. Traditionally, the UL and DL transmissions of a given two-way link are \emph{coupled}, such that they follow the same transmission path~\cite{7432156}. This has been also true for wireless systems equipped with relays, where each transmission can take either a direct or a relayed path, see Fig.~\ref{cm}. In fact, it is precisely the coupling of the relayed two-way transmission that gave rise to new schemes, such as Physical Layer Network Coding~\cite{Zhang:2006:HTP:1161089.1161129,4288792}.

Recently there have been multiple works that advocate revision of the coupled transmission dogma~\cite{7432156,6476878,6736746} in the context of heterogeneous networks (HetNets), where the uplink transmission can be made to a small cell Base Station (BS), while the downlink transmission can be made from a macro BS. The decoupling can be useful due to the fact that the transmission power of small cell BSs and macro BSs can be significantly different, but it can also be useful in terms of load balance. In~\cite{6736746}, the authors argue that the cell-centric architecture should evolve into a device-centric one, so that the connectivity and network function nearby should be tailored surrounding a specific device. The practical trial based on Vodadone's small cell test network~\cite{7037069} shows the performance gains are enough high in a dense HetNet deployment with DL and UL decoupling (DUDe). The paper~\cite{7003998} analyzes the association probabilities and average throughput of decoupled access by using stochastic geometry framework. Since the cell load and backhauling support significantly affect the association, an algorithm for cell association based on links quality, cell load and the cell backhaul capacity is proposed in~\cite{7249179}. To address the challenges in HetNets for mobile traffic offloading, best-fit cell attachment is proposed in~\cite{7248861}, where DL and UL are decoupled and attached to cells independently. In~\cite{7112544}, a tractable model is established to characterize the comprehensive SINR and rate analysis with DUDe in a multi-tier HetNets, where it is seen that decoupled association leads to significant improvement in joint UL-DL rate coverage.

The objective of this work is to bring the concept of decoupling to wireless systems that use relays. Specifically, we consider the scenario on Fig.~\ref{cm}, where multiple UEs have two-way link to the BS and the transmissions, in either UL or DL direction, can be aided by a Relay Station (RS). In this context, decoupling should be understood in the way that, for a given two-way link, the UL transmission may use e.g. the RS, while the DL transmission is made directly to the UE (or vice versa). Our model of relaying is based on a buffer-aided relaying~\cite{6330084}, where the buffer at the RS helps to take advantage of the favorable fading conditions. Buffer-aided relaying has been extensively studied in one-way scenarios, such as relay selection network~\cite{7084188}, multisource single relay network~\cite{6666121}, multisource multirelay network~\cite{6831628}. Meanwhile, bi-directional buffer-aided relaying with adaptive transmission mode selection has been investigated in delay-unconstrained case~\cite{6940315} and delay-constrained case~\cite{6940311} with fixed rate transmission; however, both works do not make use of the direct link. We are aware of~\cite{6555083} where buffer-aided relaying is applied along with direct transmission and network coding in three-node network. Since no scheduling is proposed for multiuser case, we regard~\cite{6555083} as the state-of-the-art and, as a benchmark, derive its variant for a multiuser case by adding round robin scheduling.

\begin{figure}[!t]
\centering
\subfloat[ODBA]{\includegraphics[width=2.5in]{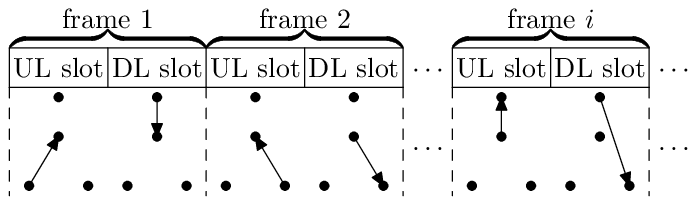}\label{odba_ex}}\linebreak
\hfil
\centering
\subfloat[NODBA]{\includegraphics[width=2.5in]{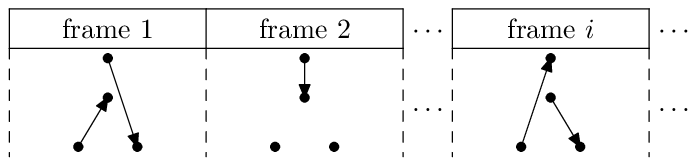}\label{nodba_ex}}\linebreak
\hfil
\centering
\subfloat[Benchmark]{\includegraphics[width=2.5in]{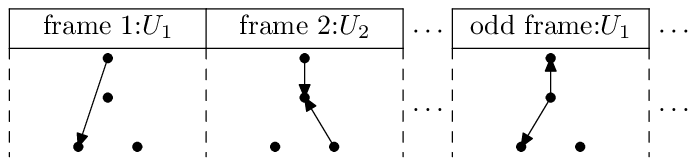}\label{ref_ex}}
\caption{Illustration of the different transmission modes.}
\label{dia_ex}
\end{figure}

In this paper, we first propose an orthogonal decoupled UL/DL buffer-aided (ODBA) relaying protocol, in which the UL selection is independent from the DL selection, based on their own instantaneous channel states and transmitter power. Next, we devise non-orthogonal decoupled UL/DL buffer-aided (NODBA) relaying protocol, where we allow opposing flows (one in UL and one in DL) to occur simultaneously and therefore interfere with each other. NODBA uses successive interference cancellation (SIC) to deal with this interference. It should be understood that ODBA and NODBA bring new building blocks that can be used to obtain complex transmission protocols. This is illustrated on Fig.~\ref{dia_ex}, where the underlying assumption is that the channel conditions change independently in each transmission frame. In the case of ODBA protocol on Fig.~\ref{odba_ex} a frame is divided into two identical slots for UL and DL traffic, respectively. The frame 2 on Fig.~\ref{odba_ex} in which the transmissions of UL and DL for the same node are identical is called a coupled frame, while the other frames are called decoupled (frames $1$ and  $i$ on Fig.~\ref{odba_ex}). NODBA protocol is shown in Fig.~\ref{nodba_ex}, where the devised non-orthogonal radio access for opposing flows is active in frame 1 and frame $i$. Finally, Fig.~\ref{ref_ex} shows a benchmark with two users, where $U_1/U_2$ is scheduled in odd/even frames and within each frame, the UE decides the transmission based on the criterion from~\cite{6555083}.

The remainder of the paper is organized as follows. In Section \ref{sys}, we present the system model. Sections \ref{sectodba} and \ref{sectnodba} present the relevant optimization problems and introduce the optimal criterion in order to achieve the maximum average two-way sum rate.  Simulation results are presented in Section \ref{simre} and the conclusions are drawn in Section \ref{conclu}.

\section{System Model}\label{sys}
The considered system consists of $M$ UEs $ U_{m},m=1,\cdots,M $, a decode-and-forward RS as $R$ and a BS as $B$. The RS on Fig.~\ref{cm} has an infinite buffer and the buffer is logically divided into two parts for storing UL and DL delay-tolerant data, respectively. Each UE has two-way traffic and each traffic source, i.e. the UEs and the BS, are backlogged and have a sufficient amount of data to send. All transceivers work in a half-duplex mode. The communication takes place in frames with fixed duration. The wireless channel is subject to block fading, the channels stay constant within a frame, but change independently from frame to frame.

We introduce the following notation for communication nodes and links. The set of nodes is denoted by ${\cal S}_{V}=\{ U_{m},R,B \}$, $\forall m \in \{ 1,\cdots,M \}$ and the transmission power of node $v$ is fixed to $P_{v} $, $v \in {\cal S}_{V}$. Without loss of generality, the power level conditions satisfies $P_{B} \geq P_{R} \geq P_{U_{m}}$. The set of UL links is ${\cal S}^{UL}= \{ U_{m}R,U_{m}B,RB \}$, while the set of DL links is  ${\cal S}^{DL}= \{  RU_{m},BU_{m},BR\} $, where $m \in \{ 1,\cdots,M \}$.

Let $Q^{UL}(i)$ and $Q^{DL}(i)$ denote the amount of normalized information in bits/symbol at the end of the $i$-th frame in the UL and DL buffer, respectively. Due to channel reciprocity, $h_{XY}(i)=h_{YX}(i)$, $XY \in {\cal S}^{DL}$,  $YX \in {\cal S}^{UL}$. 
The average channel gain of link $X^{\prime}Y^{\prime}$ is given by $\Omega_{X^{\prime}Y^{\prime}}=E\{ | h_{X^{\prime}Y^{\prime}}(i) |^2 \}$, where $X^{\prime}Y^{\prime} \in {\cal S}^{DL} \cup {\cal S}^{UL}$ and $E\{ \cdot \}$ denotes expectation. The instantaneous DL signal-to-noise ratio (SNR) with additive white Gaussian noise is given by $\gamma_{XY}^{DL} (i) \stackrel{\triangle}{=} \frac{ P_{X} | h_{XY}(i) |^2 }{ N_{0} } $, $XY \in {\cal S}^{DL} $ and corresponding instantaneous UL SNR is $\gamma_{YX}^{UL} (i) \stackrel{\triangle}{=} \frac{ P_{Y} | h_{YX}(i) |^2 }{ N_{0} } $, $YX \in {\cal S}^{UL} $.
We define:
\begin{IEEEeqnarray}{C}
C(x) \stackrel{\triangle}{=} \log_{2} (1+x) \notag
\end{IEEEeqnarray}
such that the maximal achievable transmission rate in the $i$-th frame for the link $XY$ is $C_{XY}^{DL}(i)=C(\gamma_{XY}^{DL}(i))$, $XY \in {\cal S}^{DL}$ in DL and $C_{YX}^{UL}(i)=C(\gamma_{YX}^{UL}(i))$, $YX \in {\cal S}^{UL}$ in UL. The instantaneous transmission rate of DL $XY \in {\cal  S}^{DL}$ and UL $YX \in {\cal S}^{UL}$ are given by $R_{XY}^{DL}(i)$ and $R_{YX}^{UL}(i)$.

In the case of ODBA Fig.~\ref{odba_ex}, a transmission frame is divided into an UL and DL slot, respectively. There are three possible types of transmission in an UL slot, denoted by (W1-W3): (W1) the UE $U_{m}$ transmits directly to the BS; (W2) the UE $U_{m}$ transmits to the RS; (W3) the RS transmits to the BS. Accordingly, there are three possible types of transmission in a DL slot, denoted by (W4-W6): (W4) the BS transmits to the UE $U_{m}$ directly; (W5) the RS transmits to the UE $U_{m}$; (W6) the BS transmits to the RS. The decisions on scheduling and the type of transmission are made centrally at the RS and distributed to the UEs. The CSI requirements in the ODBA protocol are (1) the BS acquires the CSI of direct DL to each UE $U_{m}$ as well as that of DL to RS; (2) each $U_{m}$ requires the CSI of the two links $U_{m}R$ and $U_{m}B$; (3) the RS needs to know the CSI of all links in order to make the decision on the selection of the transmission type that takes place in a given slot. We assume that each data transmission frame is preceded by a negligibly short procedure for CSI acquisition. Since we assume block fading, the acquired CSI is valid throughout the data transmission frame.

In NODBA protocol on Fig.~\ref{nodba_ex} UL and DL transmissions can occur simultaneously and we define four generic transmission types that can occur in a frame, denoted (T1-T4): (T1) $U_{m}$ sends to RS in the UL, while simultaneously the BS sends in the DL to another $U_{l}, l\neq m$; (T2) $U_{m}$ sends in the UL to BS directly, while simultaneously the RS sends to $U_{l}, l\neq m$; (T3) RS sends to the BS; (T4) BS sends to the RS. The types (T3) and (T4) are related to the transmissions of RS that help the data to reach its destination. On the other hand, (T1) and (T2) involve non-orthogonal transmission of opposing flows and successive interference cancellation (SIC) is applied to deal with the interference. For NODBA, both the BS and RS have to acquire the current instantaneous CSI of all the links. Similar to the case of ODBA, we assume that RS makes the decision on which transmission type should be applied in a frame and orchestrates the exchange of CSI.

The benchmark protocol on Fig.~\ref{ref_ex} is based on the transmission technique from~\cite{6555083}, supplemented with round robin scheduling. In the $i$-th frame, there are four possible types of transmission, denoted by (Z1-Z4): (Z1) the UE $U_{m}$ transmits directly to the BS; (Z2) the BS transmits directly to the UE $U_{m}$; (Z3) the UE $U_{m}$ and BS simultaneously send over a multiple access channel to the RS; (Z4) the RS broadcasts to the UE $U_{m}$ and BS, where $m \equiv i \pmod{M}$, $\bmod$ is modulo operation. The CSI acquisition and transmission strategy in each frame are according to Proposition 3 in~\cite{6555083}.

\section{Orthogonal Decoupled Uplink and Downlink Buffer-Aided Relaying Protocol}\label{sectodba}

\subsection{Instantaneous Transmission Schemes}
We use the binary variable $q_{YX}^{UL}(i)$ to indicate whether the link $YX \in {\cal S}^{UL}$ is active or silent in the UL slot of the $i$-th frame. Similarly, the binary indicator $q_{XY}^{DL}(i)$ represents the selection result for DL $XY \in {\cal S}^{DL}$ in DL slot. In each of the slots, UL or DL, only single link is active for transmission:
\begin{IEEEeqnarray}{C}
\sum_{m=1}^{M} q_{U_{m}R}^{UL}(i) + \sum_{m=1}^{M} q_{U_{m}B}^{UL}(i) +  q_{RB}^{UL}(i) =1 \notag \\
\sum_{m=1}^{M} q_{RU_{m}}^{DL}(i) + \sum_{m=1}^{M} q_{BU_{m}}^{DL}(i) +  q_{BR}^{DL}(i) =1 \notag
\end{IEEEeqnarray}
Depending on the selected transmission, the rates are determined as follows:

\noindent \textbf{(W1):} UE $U_{m}$ sends to BS directly $q_{U_{m}B}^{UL}(i)=1$. The UL transmission rate in bits per channel use is:
\begin{IEEEeqnarray}{C}
R_{U_{m}B}^{UL}(i) = \frac{1}{2} C_{U_{m}B}^{UL}(i)
\end{IEEEeqnarray}
where $\frac{1}{2}$ comes from the fact that the UL slot takes half of the frame.
The direct transmission does not affect the RS buffer:
\begin{IEEEeqnarray}{C}
Q^{UL}(i) = Q^{UL}(i-1)
\end{IEEEeqnarray}

\noindent \textbf{(W2):} UE $U_{m}$ sends to RS $q_{U_{m}R}^{UL}(i)=1$. 
The UL transmission rate in bits per channel use is: 
\begin{IEEEeqnarray}{C}
R_{U_{m}R}^{UL}(i) = \frac{1}{2} C_{U_{m}R}^{UL}(i)
\end{IEEEeqnarray}
and the change of the buffer state is:
\begin{IEEEeqnarray}{C}
Q^{UL}(i) = Q^{UL}(i-1) + R_{U_{m}R}^{UL}(i) 
\end{IEEEeqnarray}

\noindent \textbf{(W3):} RS sends to BS $q_{RB}^{UL}(i)=1$. 
The UL transmission rate in bits per channel use is: 
\begin{IEEEeqnarray}{C}
R_{RB}^{UL}(i) = \min \{ Q^{UL}(i-1), \frac{1}{2} C_{RB}^{UL}(i) \} \label{bcw3}
\end{IEEEeqnarray}
where the buffer state $Q^{UL}(i-1)$ may limit the output. The UL buffer releases space: 
\begin{IEEEeqnarray}{C}
Q^{UL}(i) = Q^{UL}(i-1) - R_{RB}^{UL}(i)
\end{IEEEeqnarray}

\noindent \textbf{(W4):} BS sends to UE $U_{m}$ directly $q_{BU_{m}}^{DL}(i)=1$.
The DL transmission rate in bits per channel use is:
\begin{IEEEeqnarray}{C}
R_{BU_{m}}^{DL}(i) = \frac{1}{2} C_{BU_{m}}^{DL}(i) 
\end{IEEEeqnarray}
The DL buffer does not change:
\begin{IEEEeqnarray}{C}
Q^{DL}(i) = Q^{DL}(i-1) 
\end{IEEEeqnarray}

\noindent \textbf{(W5):} RS sends to UE $U_{m}$ $q_{RU_{m}}^{DL}(i)=1$.
Due to the buffer, the DL transmission rate in bits per channel use is:
\begin{IEEEeqnarray}{C}
R_{RU_{m}}^{DL}(i) = \min \{ Q^{DL}(i-1), \frac{1}{2} C_{RU_{m}}^{DL}(i) \} \label{bcw5}
\end{IEEEeqnarray}
The DL buffer releases space: 
\begin{IEEEeqnarray}{C}
Q^{DL}(i) = Q^{DL}(i-1) - R_{RU_{m}}^{DL}(i) 
\end{IEEEeqnarray}

\noindent \textbf{(W6):} BS sends to RS $q_{BR}^{DL}(i)=1$. 
The DL transmission rate in bits per channel use is:
\begin{IEEEeqnarray}{C}
R_{BR}^{DL}(i) = \frac{1}{2} C_{BR}^{DL}(i)
\end{IEEEeqnarray}
The BS feeds the DL buffer:
\begin{IEEEeqnarray}{C}
Q^{DL}(i) = Q^{DL}(i-1) + R_{BR}^{DL}(i)
\end{IEEEeqnarray}

The average arrival and departure rates of the UL buffer queueing in bits per channel use are:
\begin{IEEEeqnarray}{rCl}
\bar{R}_{A}^{UL} &=& \lim_{N \to \infty} \frac{1}{N} \sum_{i=1}^{N} \sum_{m=1}^{M} 
q_{U_{m}R}^{UL}(i) R_{U_{m}R}^{UL}(i) \\
\bar{R}_{D}^{UL} &=& \lim_{N \to \infty} \frac{1}{N} \sum_{i=1}^{N} q_{RB}^{UL}(i) R_{RB}^{UL}(i)
\end{IEEEeqnarray}

Accordingly, the average arrival and departure rates of the DL buffer queueing in bits per channel use are:
\begin{IEEEeqnarray}{rCl}
\bar{R}_{A}^{DL} &=& \lim_{N \to \infty} \frac{1}{N} \sum_{i=1}^{N} q_{BR}^{DL}(i) R_{BR}^{DL}(i) \\
\bar{R}_{D}^{DL} &=& \lim_{N \to \infty} \frac{1}{N} \sum_{i=1}^{N} \sum_{m=1}^{M}  q_{RU_{m}}^{DL}(i) R_{RU_{m}}^{DL}(i)
\end{IEEEeqnarray}

As ODBA protocol is based on maximizing the UL and DL average sum rate individually, the UL buffer and DL buffer should operate at the boundary of non-absorption, which can be proved rigorously, see~\cite{6330084}. Thus the buffers should be stable and in equilibrium, we get:
\begin{IEEEeqnarray}{C}
\sum_{m=1}^{M} E \{  q_{U_{m}R}^{UL}(i) R_{U_{m}R}^{UL}(i) \} = E \{ q_{RB}^{UL}(i) R_{RB}^{UL}(i) \} \notag \\
\sum_{m=1}^{M} E \{  q_{RU_{m}}^{DL}(i) R_{RU_{m}}^{DL}(i) \} = E \{ q_{BR}^{DL}(i) R_{BR}^{DL}(i) \} \notag
\end{IEEEeqnarray}

The corresponding average sum-rate for UL and DL in bits per channel use can be expressed as:
\begin{IEEEeqnarray}{C}
\tau^{UL} = \frac{1}{N} \sum_{i=1}^{N} \Big[ \sum_{m=1}^{M} q_{U_{m}B}^{UL}(i) R_{U_{m}B}^{UL}(i) + q_{RB}^{UL}(i) R_{RB}^{UL}(i) \Big] \notag \\
\tau^{DL} = \frac{1}{N} \sum_{i=1}^{N} \Big[ \sum_{m=1}^{M} q_{BU_{m}}^{DL}(i) R_{BU_{m}}^{DL}(i) + q_{BR}^{DL}(i) R_{BR}^{DL}(i) \Big] \notag
\end{IEEEeqnarray}

\subsection{Decoupled UL and DL Optimal Transmission Strategy}
We formulate the optimization problems \textbf{P1} and \textbf{P2}:
\begin{IEEEeqnarray}{rl}
\textbf{P1:} & \quad \max_{\mathbf{q}^{UL}} \quad \tau^{UL} \label{optdbaul} \\
s.t. A1 &: \bar{R}_{A}^{UL} = \bar{R}_{D}^{UL} \notag \\
A2 &: \sum_{m=1}^{M} q_{U_{m}R}^{UL}(i) + \sum_{m=1}^{M} q_{U_{m}B}^{UL}(i) +  q_{RB}^{UL}(i) =1, \forall i \notag \\
A3 &: q_{YX}^{UL}(i) \in \{ 0,1 \}, YX \in {\cal S}^{UL}, \forall i \notag
\end{IEEEeqnarray}
\begin{IEEEeqnarray}{rl}
\textbf{P2:} & \quad \max_{\mathbf{q}^{DL}} \quad \tau^{DL} \label{optdbadl} \\
s.t. B1 &: \bar{R}_{A}^{DL} = \bar{R}_{D}^{DL} \notag \\
B2 &: \sum_{l=1}^M q_{RU_{m}}^{DL}(i) + \sum_{l=1}^M q_{BU_{m}}^{DL}(i) + q_{BR}(i) =1, \forall i \notag \\
B3 &: q_{XY}^{DL}(i) \in \{ 0,1 \}, XY \in {\cal S}^{DL}, \forall i \notag 
\end{IEEEeqnarray}
where binary indicators for UL and DL selection are collected into the decision vectors $\mathbf{q}^{UL}$ and $\mathbf{q}^{DL}$, respectively.

In the optimization problems \eqref{optdbaul} and \eqref{optdbadl}, we need to optimize the binary indicators in each frame. Note that the constraints A1 and B1 are accounted for, along with the buffer states, through \eqref{bcw3} and \eqref{bcw5}. Using the same approach as in 
~\cite{6330084,7084188,6666121,6831628,6940315,6940311,6555083}, 
we ignore the impact of the buffer state, since the event that the buffer state limits the transmission rate is negligible over a long time $N \to \infty$. This brings us to $0-1$ integer programming problems. We relax the binary constraints to the closed interval $[0,1]$, thereby enlarging the feasible solution set. However, the possible solutions of the relaxed problems lie on the boundary, and in fact they are the solutions of the original problems. The relaxed problems are solved by Lagrange multipliers and the KKT conditions.

\begin{myprop}
The optimal decision functions for maximizing the average two-way sum rate with ODBA relaying are:

UL slot Case I: If $-1 < \lambda_{1} < 0$, the criterion is
\begin{IEEEeqnarray}{rCl}
q_{U_{m}B}^{UL*}(i)
&=&
\left\{
\begin{array}{lcl}
1, & \mbox{if}  & R_{U_{m}B}^{UL}(i) \geq R_{U_{j}B}^{UL}(i), \forall j \neq m \\
   & \mbox{and} & R_{U_{m}B}^{UL}(i) \geq -\lambda_{1} R_{U_{j}R}^{UL}(i), \forall j \\
   & \mbox{and} & R_{U_{m}B}^{UL}(i) \geq (1+\lambda_{1}) R_{RB}^{UL}(i) \\
0, &            & \mbox{otherwise}
\end{array}
\right. \notag \\
q_{U_{m}R}^{UL*}(i)
&=&
\left\{
\begin{array}{lcl}
1, & \mbox{if}  & R_{U_{m}R}^{UL}(i) \geq R_{U_{j}R}^{UL}(i), \forall j \neq m \\
   & \mbox{and} & R_{U_{m}R}^{UL}(i) \geq - \frac{1}{\lambda_{1}} R_{U_{j}B}^{UL}(i), \forall j \\
   & \mbox{and} & R_{U_{m}R}^{UL}(i) \geq - \frac{ 1+\lambda_{1} }{ \lambda_{1} }   R_{RB}^{UL}(i) \\
0, &            & \mbox{otherwise}
\end{array}
\right. \notag \\
q_{RB}^{UL*}(i)
&=&
\left\{
\begin{array}{lcl}
1, & \mbox{if}  & R_{RB}^{UL}(i) \geq \frac{1}{1+\lambda_{1}} R_{U_{j}B}^{UL}(i), \forall j \\
   & \mbox{and} & R_{RB}^{UL}(i) \geq - \frac{ \lambda_{1} }{1+\lambda_{1}} R_{U_{j}R}^{UL}(i), \forall j \\
0, &            & \mbox{otherwise}
\end{array}
\right. \notag
\end{IEEEeqnarray}

UL slot Case II: If $\lambda_{1} \geq 0$ or $\lambda_{1} \leq -1$, the criterion is
\begin{IEEEeqnarray}{rCl}
q_{U_{m}B}^{UL*}(i)
&=&
\left\{
\begin{array}{lcl}
1, & \mbox{if}  & R_{U_{m}B}^{UL}(i) \geq R_{U_{j}B}^{UL}(i), \forall j \neq m \\
0, &            & \mbox{otherwise}
\end{array}
\right. \notag
\end{IEEEeqnarray}

DL slot Case I: If $-1 < \lambda_{2} < 0$, the criterion is
\begin{IEEEeqnarray}{rCl}
q_{BU_{m}}^{DL*}(i)
&=&
\left\{
\begin{array}{lcl}
1, & \mbox{if}  & R_{BU_{m}}^{DL}(i) \geq R_{BU_{j}}^{DL}(i), \forall j \neq m \\
   & \mbox{and} & R_{BU_{m}}^{DL}(i) \geq -\lambda_{2} R_{RU_{j}}^{DL}(i), \forall j \\
   & \mbox{and} & R_{BU_{m}}^{DL}(i) \geq (1+\lambda_{2}) R_{BR}^{DL}(i) \\
0, &            & \mbox{otherwise}
\end{array}
\right. \notag \\
q_{RU_{m}}^{DL*}(i)
&=&
\left\{
\begin{array}{lcl}
1, & \mbox{if}  & R_{RU_{m}}^{DL}(i) \geq R_{RU_{j}}^{DL}(i), \forall j \neq m \\
   & \mbox{and} & R_{RU_{m}}^{DL}(i) \geq - \frac{1}{\lambda_{2}} R_{BU_{j}}^{DL}(i), \forall j \\
   & \mbox{and} & R_{RU_{m}}^{DL}(i) \geq - \frac{ 1+\lambda_{2} }{ \lambda_{2} }   R_{BR}^{DL}(i) \\
0, &            & \mbox{otherwise}
\end{array}
\right. \notag \\
q_{BR}^{DL*}(i)
&=&
\left\{
\begin{array}{lcl}
1, & \mbox{if}  & R_{BR}^{DL}(i) \geq \frac{1}{1+\lambda_{2}} R_{BU_{j}}^{DL}(i), \forall j \\
   & \mbox{and} & R_{BR}^{DL}(i) \geq - \frac{ \lambda_{2} }{1+\lambda_{2}} R_{RU_{j}}^{DL}(i), \forall j \\
0, &            & \mbox{otherwise}
\end{array}
\right. \notag 
\end{IEEEeqnarray}

DL slot Case II: If $\lambda_{2} \geq 0$ or $\lambda_{2} \leq -1$, the criterion is
\begin{IEEEeqnarray}{rCl}
q_{BU_{m}}^{DL*}(i)
&=&
\left\{
\begin{array}{lcl}
1, & \mbox{if}  & R_{BU_{m}}^{DL}(i) \geq R_{BU_{j}}^{DL}(i), \forall j \neq m \\
0, &            & \mbox{otherwise}
\end{array}
\right. \notag
\end{IEEEeqnarray}
where $\lambda_{1}$ and $\lambda_{2}$ denote the Lagrange multipliers associated with the constraint A1 and B1 respectively.
\end{myprop}

\begin{IEEEproof}
Please refer to Appendix \ref{appa}.
\end{IEEEproof}

Proposition 1 specifies the optimal transmission link based on the optimal thresholds $\lambda_{1}$ and $\lambda_{2}$. Moreover, $\lambda_{1}$ and $\lambda_{2}$ are long-term dual variables which depend on the statistics of the channel and the power of the transmitters.

In UL Case I $-1 < \lambda_{1} < 0$ (or DL Case I $-1 < \lambda_{2} < 0$), $\lambda_{1}$ (or $\lambda_{2}$) under fading can be obtained numerically and iteratively with one-dimensional search using the following update equations:
\begin{IEEEeqnarray}{C}
\lambda_{1}[t+1] = \lambda_{1}[t] + \delta_{1}[t] \Delta \lambda_{1}[t] \label{update1} \\
\lambda_{2}[t+1] = \lambda_{2}[t] + \delta_{2}[t] \Delta \lambda_{2}[t] \label{update2}
\end{IEEEeqnarray}
where $t$ is the iteration index and $\delta_{1}[t], \delta_{2}[t]$ are step size which need to be chosen appropriately. In each iteration, the optimal decision vectors $\mathbf{q}^{UL*}$ and $\mathbf{q}^{DL*}$ are obtained according to Proposition 1 and then the following expressions are updated:
\begin{IEEEeqnarray}{C}
\Delta \lambda_{1}[t] = E \{ q_{RB}^{UL*}(i) R_{RB}^{UL}(i) \} - \sum_{m=1}^{M} E \{  q_{U_{m}R}^{UL*}(i) R_{U_{m}R}^{UL}(i) \} \notag \\
\Delta \lambda_{2}[t] = \sum_{m=1}^{M} E \{  q_{RU_{m}}^{DL*}(i) R_{RU_{m}}^{DL}(i) \} - E \{ q_{BR}^{DL*}(i) R_{BR}^{DL}(i) \} \notag 
\end{IEEEeqnarray}
We summarize this numerical approach in Algorithm \ref{ga}.

In UL Case II(or DL Case II), there is no need to use the RS to aid the communication. The optimal policy is just to select the maximal direct link transmission.

\begin{figure}[!t]
\begin{algorithm}[H]
\caption{1D Search for $\lambda_{1}^{*}$ and $\lambda_{2}^{*}$ respectively}\label{ga}
\begin{algorithmic}[1]
\State \textbf{initialize} $t=0$, $\lambda_{1}[0]$ and $\lambda_{2}[0]$
\Repeat
	\State Compute $\mathbf{q}^{UL*}$ and $\mathbf{q}^{DL*}$ according to Proposition 1
	\State Compute $\Delta \lambda_{1}[t]$ and $\Delta \lambda_{2}[t]$
	\State Update $\lambda_{1}[t+1]$ and $\lambda_{2}[t+1]$ based on \eqref{update1} and \eqref{update2}
	\State $t \gets t+1$
\Until{converge to $\lambda_{1}^{*}$ and $\lambda_{2}^{*}$}
\end{algorithmic}
\end{algorithm}
\end{figure}

\section{Non-orthogonal Decoupled Uplink and Downlink Buffer-Aided Relaying Protocol}\label{sectnodba}

\subsection{Instantaneous Transmission Schemes}
In both transmission types (T1) or (T2), two links are active simultaneously. We use the binary variables $q_{(m,l)}^{T1}(i)$ and $q_{(m,l)}^{T2}(i)$  to indicate whether the transmission type (T1) and (T2) takes place, respectively, in the $i$-th frame. $q_{RB}^{T3}(i)$ and $q_{BR}^{T4}(i)$ are used to indicate the transmission type (T3) and (T4). There are $M(M-1)$ possibilities to make a transmission of type (T1) or (T2). In each frame only one of the possible transmission types takes place, such that:
\begin{IEEEeqnarray}{C}
\sum_{m} \sum_{ l \neq m} q_{(m,l)}^{T1}(i) + \sum_{m} \sum_{ l \neq m} q_{(m,l)}^{T2}(i) + q_{RB}^{T3}(i) +q_{BR}^{T4}(i) = 1 \notag
\end{IEEEeqnarray}
Depending on the selected transmission types, the rates are determined as follows:

\noindent \textbf{(T1):} UE $U_{m}$ sends to RS in the UL and simultaneously BS sends to another UE $U_{l}$, $l \neq m$ in the DL $q_{(m,l)}^{T1}(i)=1$.
Although the interference occurs at the RS, with the help of SIC, the UL rate could achieve its capacity $C_{U_{m}R}^{UL}(i)$ by limiting the DL rate:
\begin{IEEEeqnarray}{C}
R_{BU_{l}}^{T1}(i) \leq C( \frac{ \gamma_{BR}^{DL}(i) }{ 1+\gamma_{U_{m}R}^{UL}(i) } )  \label{bound1}
\end{IEEEeqnarray}
On the other hand, the direct DL rate is limited by:
\begin{IEEEeqnarray}{C}
R_{BU_{l}}^{T1}(i) \leq C_{BU_{l}}^{DL}(i) \label{bound2}
\end{IEEEeqnarray}
Hence the instantaneous UL/DL transmission rates in bits per channel use are determined as:
\begin{IEEEeqnarray}{rCl}
R_{BU_{l}}^{T1}(i) &=&  \min \{  C_{BU_{l}}^{DL}(i) , C( \frac{ \gamma_{BR}^{DL}(i) }{ 1+\gamma_{U_{m}R}^{UL}(i) } ) \}  \label{ndbatype1} \\
R_{U_{m}R}^{T1}(i) &=& C_{U_{m}R}^{UL}(i)
\end{IEEEeqnarray}
where the direct DL rate is decreased to the minor bound of \eqref{bound1} and \eqref{bound2}. The UL buffer updates as:
\begin{IEEEeqnarray}{C}
Q^{UL}(i)=Q^{UL}(i-1) + R_{U_{m}R}^{T1}(i)
\end{IEEEeqnarray}

\noindent \textbf{(T2):} UE $U_{m}$ sends in the UL to BS directly and simultaneously RS sends to another UE $U_{l}$, $l \neq m$ in the DL $q_{(m,l)}^{T2}(i)=1$.
Similarly, SIC is applied to deal with the interference at the BS, the instantaneous UL/DL transmission rates in bits per channel use are determined as:
\begin{align}
R_{RU_{l}}^{T2}(i) &=  \min \{ Q^{DL}(i-1), C_{RU_{l}}^{DL}(i), C( \frac{ \gamma_{RB}^{UL}(i) }{ 1+\gamma_{U_{m}B}^{UL}(i) } )  \} \label{ndbatype2} \\
R_{U_{m}B}^{T2}(i) &= C_{U_{m}B}^{UL}(i)
\end{align}
where the DL buffer accumulation $Q^{DL}(i-1)$ may also limit the output. The DL buffer releases space: 
\begin{IEEEeqnarray}{C}
Q^{DL}(i)=Q^{DL}(i-1) - R_{RU_{l}}^{T2}(i)
\end{IEEEeqnarray}

\noindent \textbf{(T3):} RS sends to BS $q_{RB}^{T3}(i)=1$ with the transmission rate: 
\begin{IEEEeqnarray}{C}
R_{RB}^{T3}(i) = \min \{ Q^{UL}(i-1),C_{RB}^{UL}(i) \}
\end{IEEEeqnarray}
The UL buffer releases space: 
\begin{IEEEeqnarray}{C}
Q^{UL}(i)=Q^{UL}(i-1) - R_{RB}^{T3}(i)
\end{IEEEeqnarray}

\noindent \textbf{(T4):} BS sends to RS $q_{BR}^{T4}(i)=1$ with the transmission rate: 
\begin{IEEEeqnarray}{C}
R_{BR}^{T4}(i) = C_{BR}^{DL}(i) 
\end{IEEEeqnarray}
The DL buffer updates as: 
\begin{IEEEeqnarray}{C}
Q^{DL}(i)=Q^{DL}(i-1) + R_{BR}^{T4}(i)
\end{IEEEeqnarray}

The average arrival and departure rates of the UL buffer queueing in bits per channel use are: 
\begin{IEEEeqnarray}{rCl}
\bar{R}_{A}^{UL} &=& \lim_{N \to \infty} \frac{1}{N} \sum_{i=1}^{N} \sum_{m}
\sum_{l \neq m} q_{(m,l)}^{T1}(i) R_{U_{m}R}^{T1}(i) \\
\bar{R}_{D}^{UL} &=& \lim_{N \to \infty} \frac{1}{N} \sum_{i=1}^{N} q_{RB}^{T3}(i) R_{RB}^{T3}(i)
\end{IEEEeqnarray}

Accordingly, the average arrival and departure rates of the DL buffer queueing in bits per channel use are:
\begin{IEEEeqnarray}{rCl}
\bar{R}_{A}^{DL} &=& \lim_{N \to \infty} \frac{1}{N} \sum_{i=1}^{N} q_{BR}^{T4}(i) R_{BR}^{T4}(i) \\
\bar{R}_{D}^{DL} &=& \lim_{N \to \infty} \frac{1}{N} \sum_{i=1}^{N} \sum_{m} \sum_{l \neq m} q_{(m,l)}^{T2}(i) R_{RU_{l}}^{T2}(i) 
\end{IEEEeqnarray}

Similarly, equilibrium of the UL buffer and DL buffer should be maintained, which lead to the following expressions:
\begin{IEEEeqnarray}{rCl}
\sum_{m} \sum_{l \neq m} E \{  q_{(m,l)}^{T1}(i) R_{U_{m}R}^{T1}(i) \} &=& E \{ q_{RB}^{T3}(i) R_{RB}^{T3}(i) \} \notag \\
\sum_{m} \sum_{l \neq m} E \{  q_{(m,l)}^{T2}(i) R_{RU_{l}}^{T2}(i) \} &=& E \{ q_{BR}^{T4}(i) R_{BR}^{T4}(i) \} \notag
\end{IEEEeqnarray}
with the corresponding average two-way sum rate: 
\begin{IEEEeqnarray}{C}
\tau = \frac{1}{N} \sum_{i=1}^{N} \Big[  
 \sum_{m} \sum_{l \neq m} q_{(m,l)}^{T1}(i) R_{BU_{l}}^{T1}(i) + q_{RB}^{T3}(i) R_{RB}^{T3}(i) + \sum_{m} \sum_{l \neq m} q_{(m,l)}^{T2}(i) \big[ R_{U_{m}B}^{T2}(i) + R_{RU_{l}}^{T2}(i)   \big] \Big] \notag
\end{IEEEeqnarray}

\subsection{Decoupled UL and DL Optimal Transmission Strategy}
We formulate the optimization problem as \textbf{P3}:
\begin{IEEEeqnarray}{rl}
\textbf{P3:} & \quad \max_{\mathbf{q}} \quad \tau \label{optndba} \\
s.t. C1: & \bar{R}_{A}^{UL} = \bar{R}_{D}^{UL} \notag \\
C2: & \bar{R}_{A}^{DL} = \bar{R}_{D}^{DL} \notag \\
C3: & \sum_{m} \sum_{ l \neq m} q_{(m,l)}^{T1}(i) + \sum_{m} \sum_{l \neq m} q_{(m,l)}^{T2}(i) + q_{RB}^{T3}(i) +q_{BR}^{T4}(i) = 1, \forall i \notag \\
C4: & q_{(m,l)}^{T1}(i), q_{(m,l)}^{T2}(i), q_{RB}^{T3}(i), q_{BR}^{T4}(i) \in \{ 0,1 \},  \forall m, \forall l \neq m, \forall i \notag
\end{IEEEeqnarray}  
where decision indicators are collected in the vector $\mathbf{q}$. We solve \textbf{P3} similar to \textbf{P1} and \textbf{P2} and put forward \emph{Proposition 2}.

\begin{myprop}
The optimal decision functions for maximizing the average two-way sum rate with NODBA relaying protocol are:

Case I: If $\lambda_{3} > -1$ and $\lambda_{4} < 0$, the criterion is
\begin{IEEEeqnarray}{rCl}
q_{(m,l)}^{T1*}(i)
&=&
\left\{
\begin{array}{lcl}
1, & \mbox{if}  & \Lambda_{(m,l)}^{T1}(i) \geq \Lambda_{(j,k)}^{T1}(i), \\
   &            & \forall (j,k) \neq (m,l) \\
   & \mbox{and} & \Lambda_{(m,l)}^{T1}(i) \geq \Lambda_{(j,k)}^{T2}(i), \forall (j,k) \\
   & \mbox{and} & \Lambda_{(m,l)}^{T1}(i) \geq \Lambda_{RB}^{T3}(i) \\
   & \mbox{and} & \Lambda_{(m,l)}^{T1}(i) \geq \Lambda_{BR}^{T4}(i) \\
0, &            & \mbox{otherwise}
\end{array}
\right. \notag \\
q_{(m,l)}^{T2*}(i)
&=&
\left\{
\begin{array}{lcl}
1, & \mbox{if}  & \Lambda_{(m,l)}^{T2}(i) \geq \Lambda_{(j,k)}^{T2}(i), \\
   &            & \forall (j,k) \neq (m,l) \\
   & \mbox{and} & \Lambda_{(m,l)}^{T2}(i) \geq \Lambda_{(j,k)}^{T1}(i), \forall (j,k) \\
   & \mbox{and} & \Lambda_{(m,l)}^{T2}(i) \geq \Lambda_{RB}^{T3}(i) \\
   & \mbox{and} & \Lambda_{(m,l)}^{T2}(i) \geq \Lambda_{BR}^{T4}(i) \\
0, &            & \mbox{otherwise}
\end{array}
\right. \notag \\
q_{RB}^{T3*}(i)
&=&
\left\{
\begin{array}{lcl}
1, & \mbox{if}  & \Lambda_{RB}^{T3}(i) \geq \Lambda_{(j,k)}^{T1}(i), \forall (j,k) \\
   & \mbox{and} & \Lambda_{RB}^{T3}(i) \geq \Lambda_{(j,k)}^{T2}(i), \forall (j,k) \\
   & \mbox{and} & \Lambda_{RB}^{T3}(i) \geq \Lambda_{BR}^{T4}(i) \\
0, &            & \mbox{otherwise}
\end{array}
\right. \notag \\
q_{BR}^{T4*}(i)
&=&
\left\{
\begin{array}{lcl}
1, & \mbox{if}  & \Lambda_{BR}^{T4}(i) \geq \Lambda_{(j,k)}^{T1}(i), \forall (j,k) \\
   & \mbox{and} & \Lambda_{BR}^{T4}(i) \geq \Lambda_{(j,k)}^{T2}(i), \forall (j,k) \\
   & \mbox{and} & \Lambda_{BR}^{T4}(i) \geq \Lambda_{RB}^{T3}(i) \\
0, &            & \mbox{otherwise}
\end{array}
\right. \notag
\end{IEEEeqnarray}

Case II: If $\lambda_{3} > -1$ and $\lambda_{4} \geq 0$, the criterion is
\begin{IEEEeqnarray}{rCl}
q_{(m,l)}^{T1*}(i)
&=&
\left\{
\begin{array}{lcl}
1, & \mbox{if}  & \Lambda_{(m,l)}^{T1}(i) \geq \Lambda_{(j,k)}^{T1}(i), \\
   &            & \forall (j,k) \neq (m,l) \\
   & \mbox{and} & \Lambda_{(m,l)}^{T1}(i) \geq \Lambda_{RB}^{T3}(i) \\
0, &            & \mbox{otherwise}
\end{array}
\right. \notag \\
q_{RB}^{T3*}(i)
&=&
\left\{
\begin{array}{lcl}
1, & \mbox{if}  & \Lambda_{RB}^{T3}(i) \geq \Lambda_{(j,k)}^{T1}(i), \forall (j,k) \\
0, &            & \mbox{otherwise}
\end{array}
\right. \notag
\end{IEEEeqnarray}

Case III: If $\lambda_{3} \leq -1$ and $\lambda_{4} < 0$, the criterion is
\begin{IEEEeqnarray}{rCl}
q_{(m,l)}^{T2*}(i)
&=&
\left\{
\begin{array}{lcl}
1, & \mbox{if}  & \Lambda_{(m,l)}^{T2}(i) \geq \Lambda_{(j,k)}^{T2}(i), \\
   &            & \forall (j,k) \neq (m,l) \\
   & \mbox{and} & \Lambda_{(m,l)}^{T2}(i) \geq \Lambda_{BR}^{T4}(i) \\
0, &            & \mbox{otherwise}
\end{array}
\right. \notag \\
q_{BR}^{T4*}(i)
&=&
\left\{
\begin{array}{lcl}
1, & \mbox{if} & \Lambda_{BR}^{T4}(i) \geq \Lambda_{(j,k)}^{T2}(i), \forall (j,k) \\
0, &            & \mbox{otherwise}
\end{array}
\right. \notag
\end{IEEEeqnarray}
where selection matrices are denoted by
\begin{IEEEeqnarray}{rCl}
\Lambda_{(x,y)}^{T1}(i) & = & R_{BU_{y}}^{T1}(i) - \lambda_{3} R_{U_{x}R}^{T1}(i) \notag \\
\Lambda_{(x,y)}^{T2}(i) & = & R_{U_{x}B}^{T2}(i) + (1+ \lambda_{4}) R_{RU_{y}}^{T2}(i) \notag \\
\Lambda_{RB}^{T3}(i)    & = & (1+\lambda_{3}) R_{RB}^{T3}(i) \notag \\
\Lambda_{BR}^{T4}(i)    & = & - \lambda_{4}  R_{BR}^{T4}(i) \notag
\end{IEEEeqnarray}
and $\lambda_{3}, \lambda_{4}$ denote the Lagrange multiplier corresponding to constraint C1 and C2. 
\end{myprop}

\begin{IEEEproof}
Please refer to Appendix \ref{appb}.
\end{IEEEproof}

Since long-term dual variables $\lambda_{3}$ and $\lambda_{4}$ are not independent in NODBA, we adopt two-dimensional search to find the optimal thresholds using the following update equation:
\begin{IEEEeqnarray}{C}
\lambda_{3}[t+1] = \lambda_{3}[t] + \delta_{3}[t] \Delta \lambda_{3}[t] \label{update3} \\
\lambda_{4}[t+1] = \lambda_{4}[t] + \delta_{4}[t] \Delta \lambda_{4}[t] \label{update4}
\end{IEEEeqnarray}
where $t$ is the iteration index and $\delta_{3}[t], \delta_{4}[t]$ are step size. Moreover, the optimal decision vector $\mathbf{q}^{*}$ is updated in each iteration according to Proposition 2 along with the following expressions:
{\small
\begin{IEEEeqnarray}{C}
\Delta \lambda_{3} [t] =  E \{ q_{RB}^{T3*}(i) R_{RB}^{T3}(i) \} - \sum_{m} \sum_{l \neq m} E \{  q_{(m,l)}^{T1*}(i) R_{U_{m}R}^{T1}(i) \} \notag \\
\Delta \lambda_{4} [t] =  \sum_{m} \sum_{l \neq m} E \{  q_{(m,l)}^{T2*}(i) R_{RU_{l}}^{T2}(i) \} -  E \{ q_{BR}^{T4*}(i) R_{BR}^{T4}(i) \} \notag 
\end{IEEEeqnarray}
}
We summarize 2D search in Algorithm \ref{2dga}.

Case I indicates that both non-orthogonal transmission types (T1) and (T2) offer benefits over a long term. On the other hand, in Case II and Case III, only one of the devised non-orthogonal transmission types, either (T1) or (T2), has the potential to improve the average two-way sum rate.

\begin{figure}[!t]
\begin{algorithm}[H]
\caption{2D search for $\lambda_{3}^{*}$ and $\lambda_{4}^{*}$ }
\label{2dga}
\begin{algorithmic}[1]
\State \textbf{initialize} $t=0$ and $\lambda_{3}[0]$,$\lambda_{4}[0]$
\Repeat
	\State Compute $\mathbf{q}^{*}$ according to Proposition 2
	\State Compute $\Delta \lambda_{3}[t]$ and $\Delta \lambda_{4}[t]$
	\State Update $\lambda_{3}[t+1]$ and $\lambda_{4}[t+1]$ based on \eqref{update3} and \eqref{update4}
	\State $t \gets t+1$
\Until{converge to $\lambda_{3}^{*}$ and $\lambda_{4}^{*}$ }
\end{algorithmic}
\end{algorithm}
\end{figure}

\section{Simulation Results}\label{simre}
We present simulation results to compare the performance of the proposed ODBA and NODBA protocols with state-of-the-art scheme from~\cite{6555083}, which is here combined with a round robin scheduler. The evaluation scenario has 2 UEs.  All links are subject to Rayleigh fading. We denote the average channel gain vector of all the involved links $\mathbf{\Omega} =  [ \Omega_{U_{1}R}, \Omega_{U_{2}R}, \Omega_{U_{1}B}, \Omega_{U_{2}B}, \Omega_{RB} ]$. The noise power is normalized to $1$.

\begin{figure}[!t]
\centering
\includegraphics[width=2.5in]{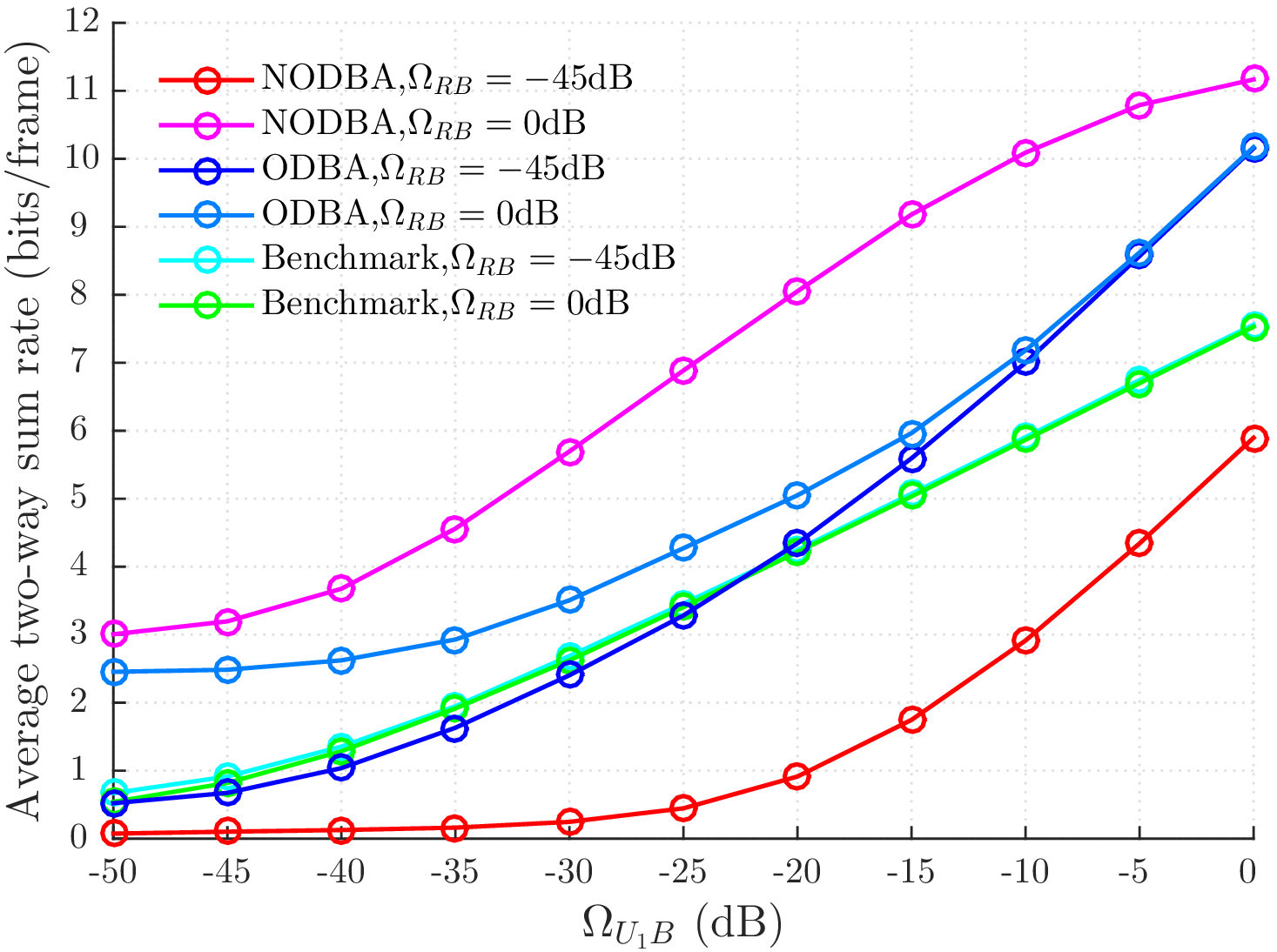}
\caption{Average two-way sum rate vs. $\Omega_{U_{1}B}$ for $P_{U_{1}}=P_{U_{2}}=P_{R}=20$dBm, $P_{B}=46$dBm and $\mathbf{\Omega}=[-13,-12,\Omega_{U_{1}B},-49,\Omega_{RB}]$dB.}
\label{liurk3}
\end{figure}

Fig.~\ref{liurk3} depicts the average two-way sum rate as a function of $\Omega_{U_{1}B}$. We consider two situations with respect to the link $RB$: $\Omega_{RB}=-45$dB and $\Omega_{RB}=0$dB. When the RS-to-BS link is good $\Omega_{RB}=0$dB, both ODBA and NODBA outperform the benchmark. On the other hand, when the RS-to-BS link is a bottleneck with $\Omega_{RB}=-45$dB, ODBA exceed the benchmark when $\Omega_{U_{1}B} \geq -22$dB, while NODBA is inferior to the benchmark since it relies on SIC. We should note that ODBA takes the advantage of multiuser diversity, while for the benchmark, the performance deterioration in even frames for $U_{2}$ could not be improved by increasing $\Omega_{U_{1}B}$.

\begin{figure}[!t]
\centering
\includegraphics[width=2.5in]{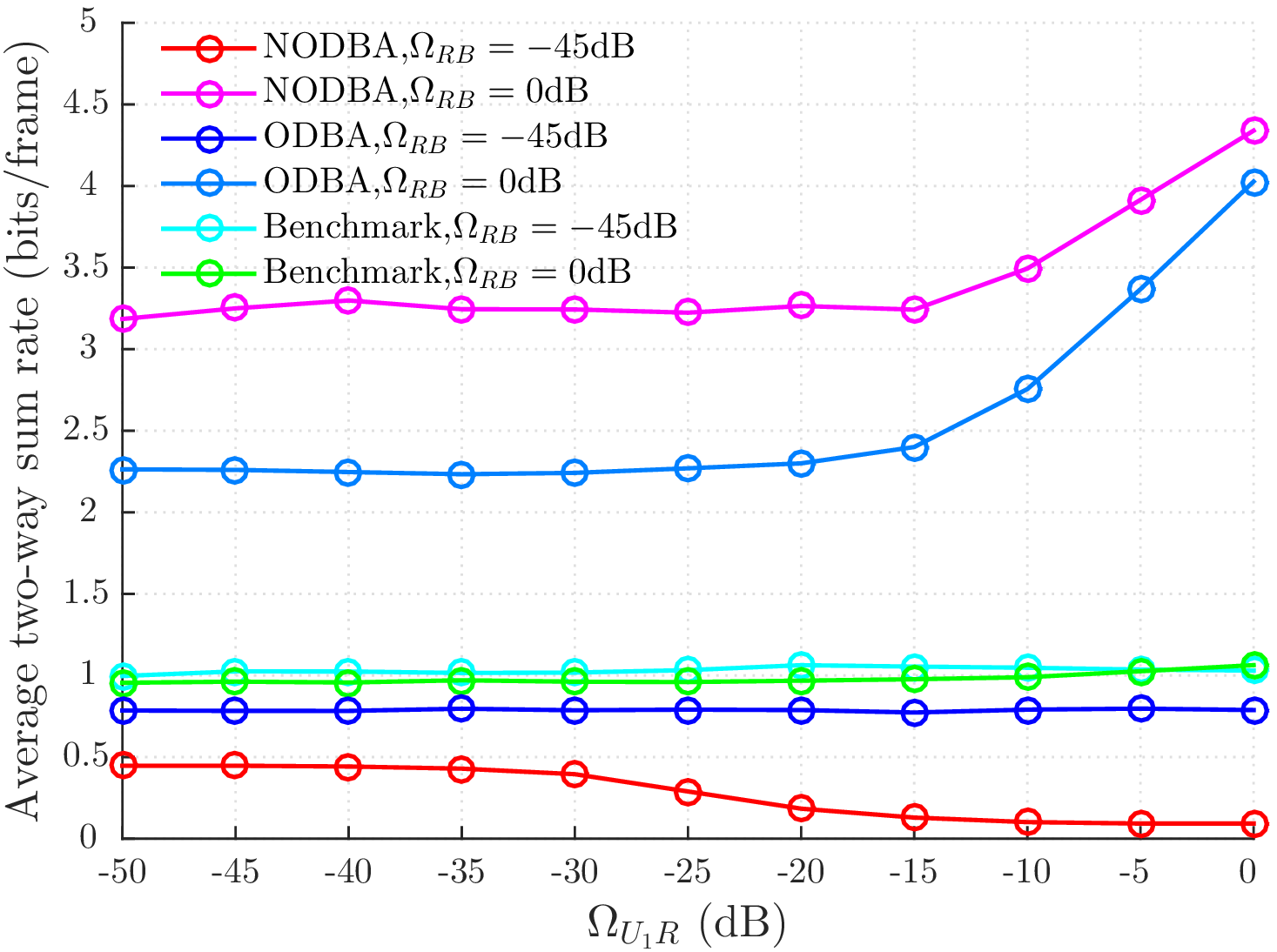}
\caption{Average two-way sum rate vs. $\Omega_{U_{1}R}$ for $P_{U_{1}}=P_{U_{2}}=P_{R}=20$dBm, $P_{B}=46$dBm and $\mathbf{\Omega}=[\Omega_{U_{1}R},-12,-43,-49,\Omega_{RB}]$dB.}
\label{liurk4}
\end{figure}

Fig.~\ref{liurk4} shows the average two-way sum rate as a function of $\Omega_{U_{1}R}$. When $\Omega_{U_{1}R}<-15$dB, the two-way sum rate grows very slowly, since in most of the attempts the direct link is selected for transmission, even though the direct links are statistically weaker than the relayed links. When $\Omega_{U_{1}R}>-15$dB, ODBA and NODBA display their superiority with a strong RS-to-BS link. Again, the performance of the proposed protocols is severely degraded when the RS-to-BS link is a bottleneck.

\begin{figure}[!t]
\centering
\includegraphics[width=2.5in]{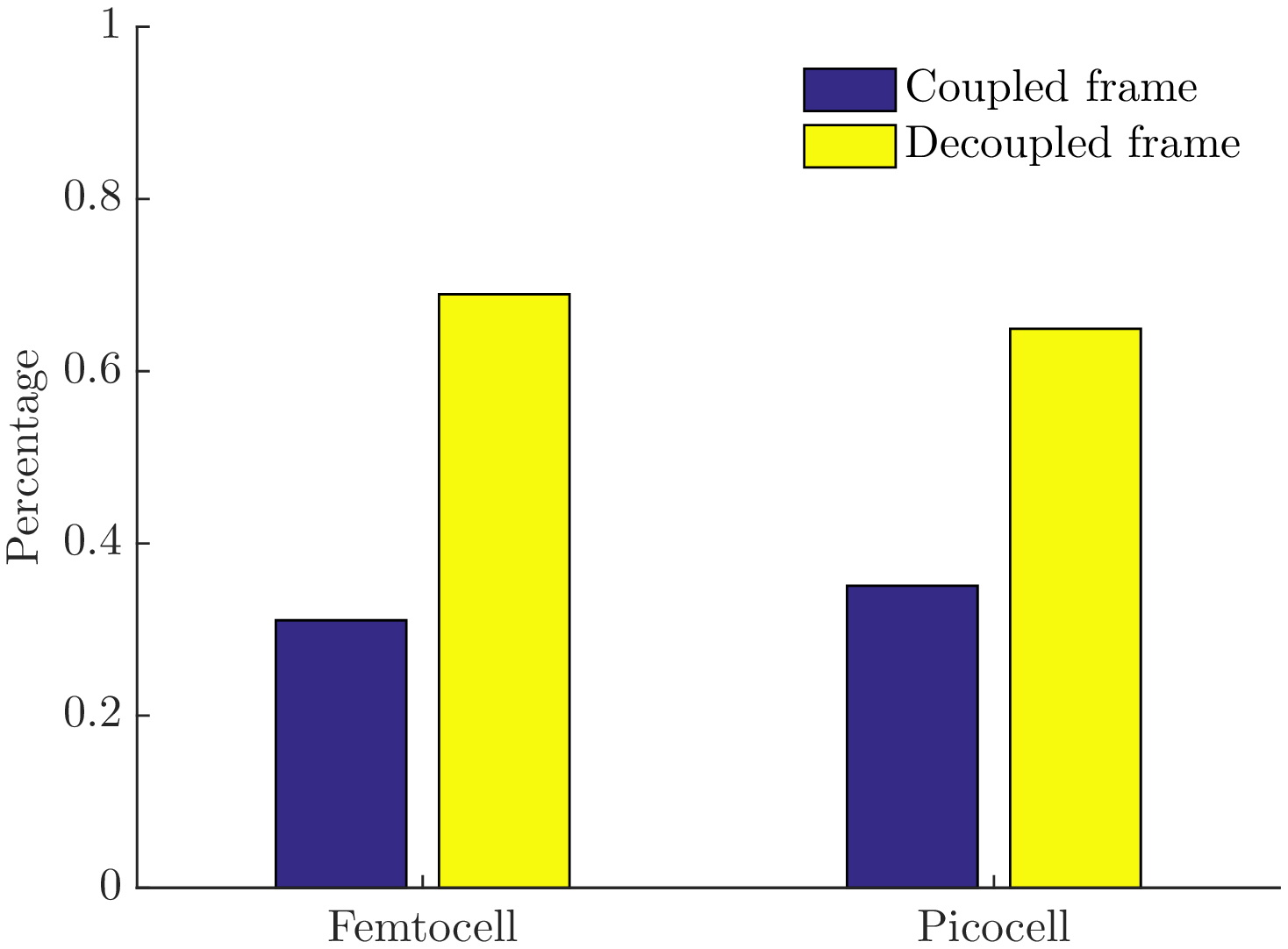}
\caption{Percentage of coupled frame and decoupled frame with ODBA protocol for $\mathbf{\Omega}=[-6,-8,-40,-41,0]$dB and $P_{U_{1}}=P_{U_{2}}=20$dBm, $P_{B}=46$dBm (1) Femtocell $P_{R}=20$dBm; (2) Picocell $P_{R}=30$dBm.}
\label{liurk5}
\end{figure}

\begin{figure}[!t]
\centering
\includegraphics[width=2.5in]{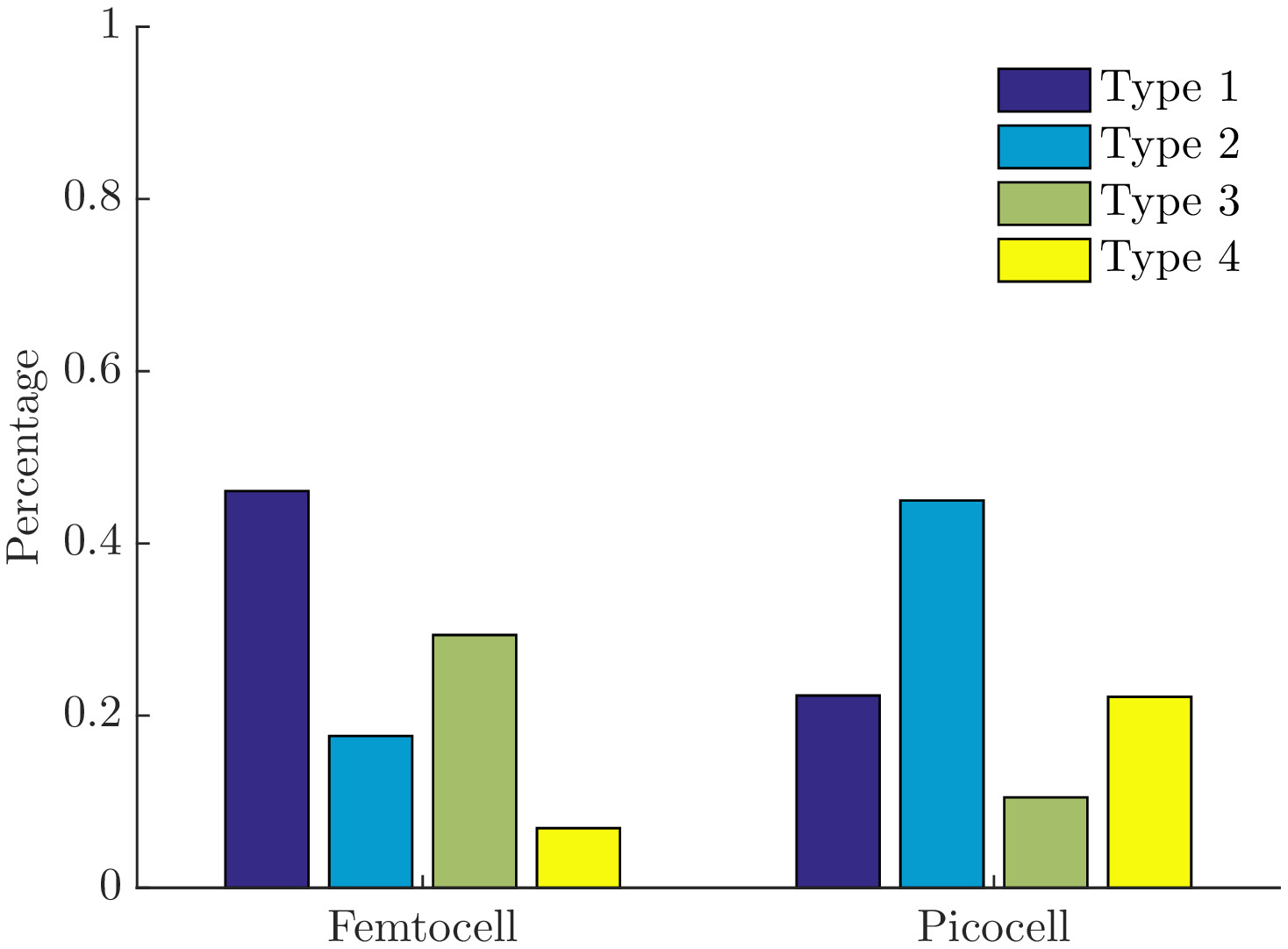}
\caption{Percentage of different transmission types with NODBA protocol for $\mathbf{\Omega}=[-6,-8,-40,-41,0]$dB and $P_{U_{1}}=P_{U_{2}}=20$dBm, $P_{B}=46$dBm (1) Femtocell $P_{R}=20$dBm; (2) Picocell $P_{R}=30$dBm.}
\label{liurk6}
\end{figure}

Fig.~\ref{liurk5} shows the percentage of frames in which the ODBA protocol selects coupled and decoupled transmissions, respectively, when the RS-to-BS link is strong. This is shown for two cases, each corresponding to a different power level used by the RS: (1) Femtocell level and (2) Picocell level. It is interesting to see that decoupled transmission is selected more often, which further justifies its introduction. Fig.~\ref{liurk6} shows a similar type of statistics for the NODBA protocol, depicting how often each transmission type is selected. When the RS power is at a femtocell level, transmission type (T1) dominates, while when the RS power is at a picocell level, transmission type (T2) dominates.

\begin{figure*}[!t]
\centering
\subfloat[ODBA]{\includegraphics[width=2.5in]{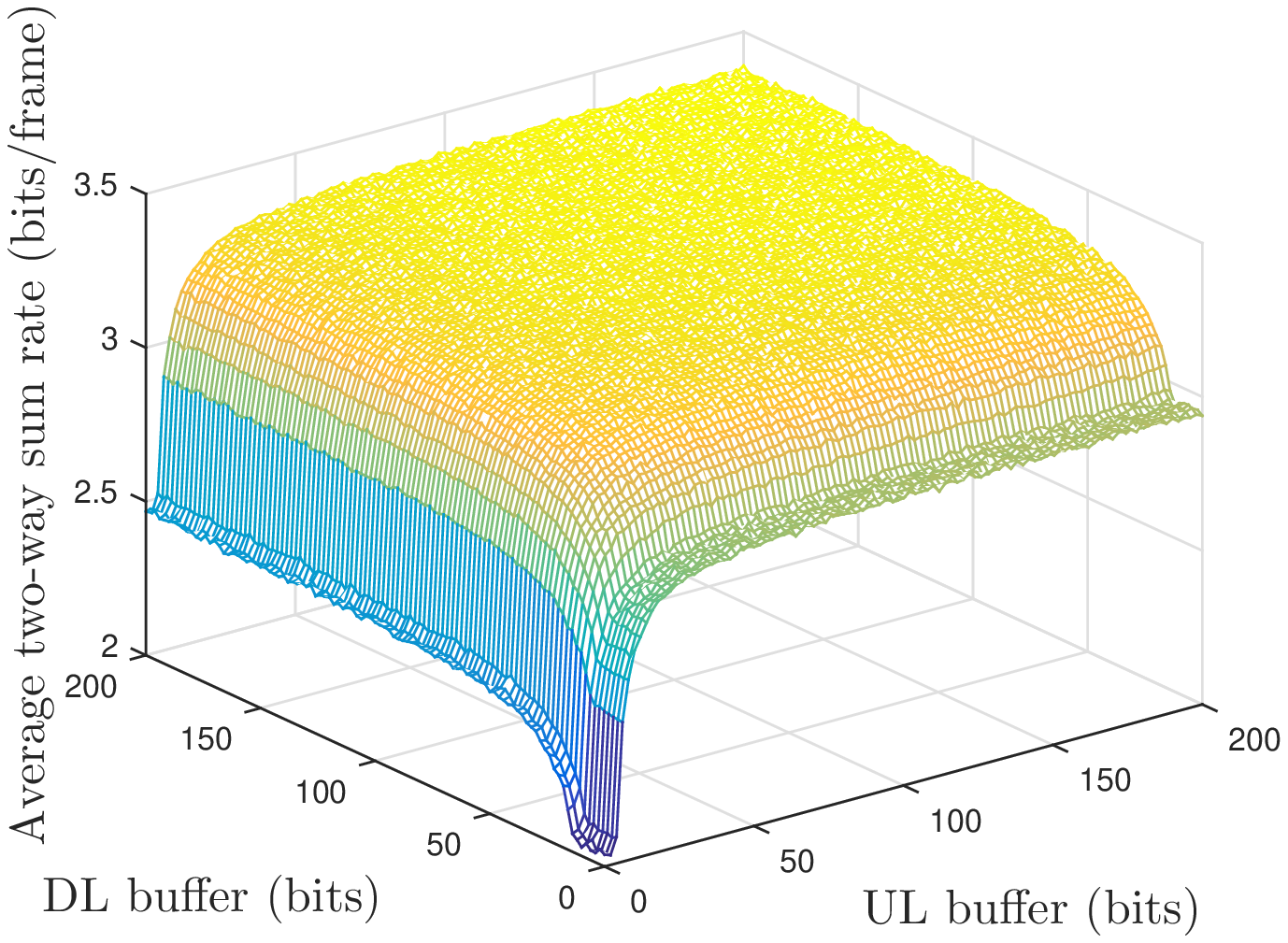}
\label{liurk7a}}
\hfil
\subfloat[NODBA]{\includegraphics[width=2.5in]{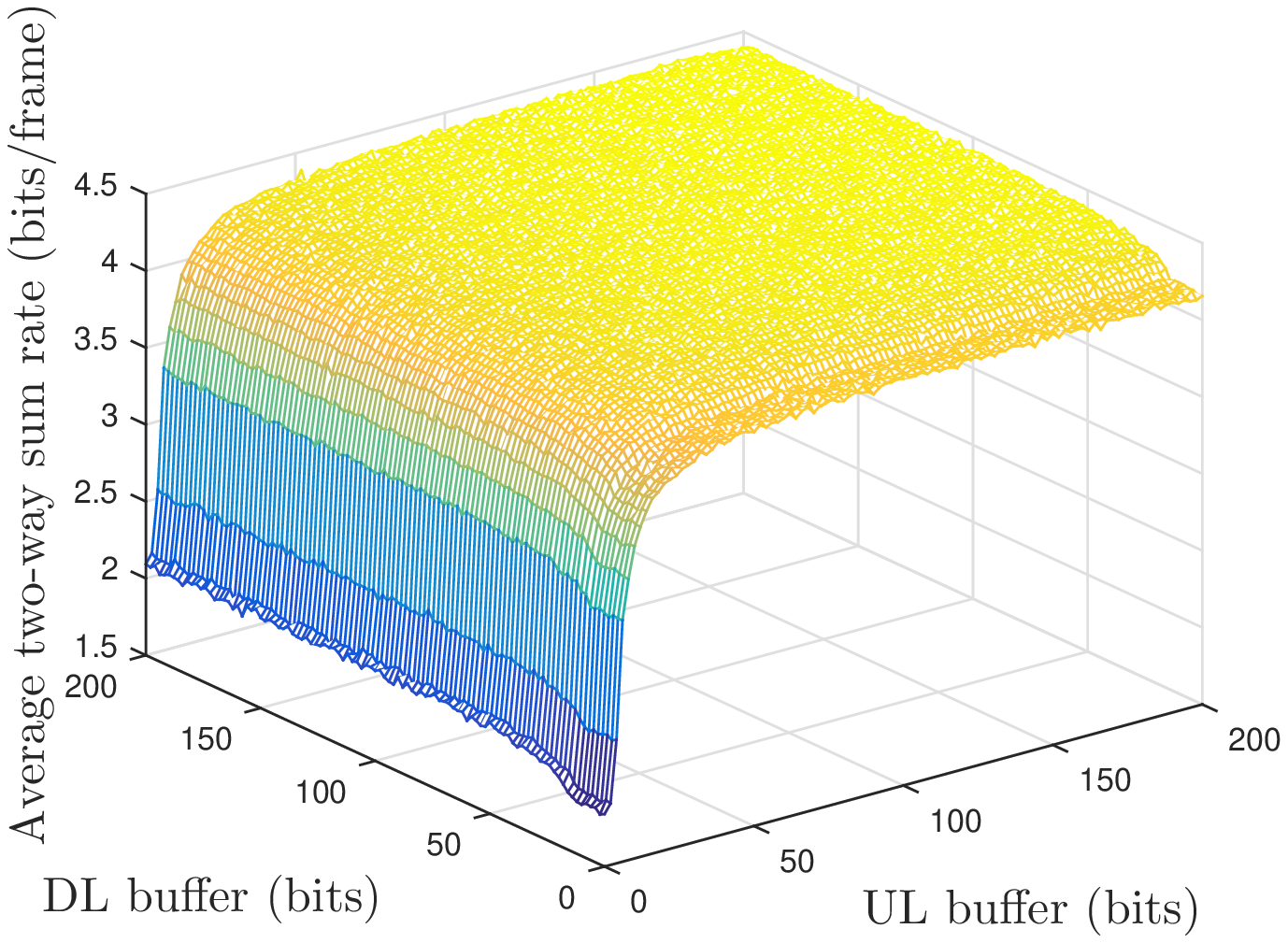}
\label{liurk7b}}
\caption{Average two-way sum rate vs. UL and DL buffer size for $\mathbf{\Omega}=[-6,-8,-40,-41,0]$dB and $P_{U_{1}}=P_{U_{2}}=P_{R}=20$dBm, $P_{B}=46$dBm.}
\label{liurk7}
\end{figure*}

In Fig.~\ref{liurk7}, we investigate the effect of the UL and DL buffer size. In practice, a buffer is finite and it is necessary to avoid overflow. We therefore heuristically modify the policy in Proposition 1 and Proposition 2 in the following way. We put a constraint that the relay never overflows, in both UL and DL, such that the UE (BS) feeds the buffer only if it has sufficient space.Additionally, the UL and DL buffer provide output bits only if the buffer is not empty. Clearly, the modifications worsen the performance of the scheme that is designed under the assumption of infinite buffers, but one can find appropriate UL and DL buffer size in order to approximate the optimal performance.

\section{Conclusion}\label{conclu}
Decoupling the path of uplink (UL) and downlink (DL) transmission for two-way links opens up new design possibilities in wireless networks. In this paper we consider a scenario in which the two-way link is between a UE (User Equipment) and a Base Station (BS). The communication in each direction can be aided by a buffer-equipped Relay Station (RS). In this context, decoupling of a two-way link means that one of the UL/DL directions uses direct transmission between the UE and the BS, while the other direction is relayed through the RS. We propose 
two protocols that make use of decoupled transmission: orthogonal decoupled UL/DL buffer-aided (ODBA) relaying protocol and non-orthogonal decoupled UL/DL buffer-aided (NODBA) relaying protocol. In ODBA, mutual independent selections take place for UL and DL transmission, while in NODBA, UL and DL are active simultaneously and the receiver uses Successive Interference Cancellation (SIC). We derive the optimal criterion based on the average channel gain and instantaneous CSI of the involved links. The numerical results show that decoupling can bring advantages in terms of average two-way sum rate, in particular when the RS-BS link is strong.

\appendices
\section{}\label{appa}
Here we briefly show how to solve the relaxed optimization problem in ODBA protocol. The Lagrangian functions of the relaxed problems with KKT conditions for UL and DL are given by

\begin{IEEEeqnarray}{l}
{\cal L}^{UL}= - \frac{1}{N} \sum_{i=1}^{N} \Big[ \sum_{m=1}^{M} q_{U_{m}B}^{UL}(i) R_{U_{m}B}^{UL}(i) + q_{RB}^{UL}(i) R_{RB}^{UL}(i) \Big] \notag \\
+ \lambda_{1} \frac{1}{N} \sum_{i=1}^{N} \Big[ \sum_{m=1}^{M} q_{U_{m}R}^{UL}(i) R_{U_{m}R}^{UL}(i) -  q_{RB}^{UL}(i) R_{RB}^{UL}(i) \Big] \notag \\
+ \sum_{i=1}^{N} \alpha^{UL} (i) \Big[ \sum_{m=1}^{M} q_{U_{m}R}^{UL}(i) + \sum_{m=1}^{M} q_{U_{m}B}^{UL}(i) +  q_{RB}^{UL}(i) -1 \Big] \notag \\
+  \sum_{i=1}^{N} \sum_{YX} \eta_{YX}^{UL}(i) \Big[ q_{YX}^{UL}(i)-1 \Big] - \sum_{i=1}^{N} \sum_{YX} \xi_{YX}^{UL}(i) q_{YX}^{UL}(i) \notag
\end{IEEEeqnarray}
where $\lambda_{1} , \alpha^{UL} (i), \eta_{YX}^{UL}(i), \xi_{YX}^{UL}(i)$ are Lagrange multipliers.

\begin{IEEEeqnarray}{l}
{\cal L}^{DL}= - \frac{1}{N} \sum_{i=1}^{N} \Big[ \sum_{m=1}^{M} q_{BU_{m}}^{DL}(i) R_{BU_{m}}^{DL}(i) + q_{BR}^{DL}(i) R_{BR}^{DL}(i) \Big] \notag \\ 
+ \lambda_{2} \frac{1}{N} \sum_{i=1}^{N} \Big[ \sum_{m=1}^{M} q_{RU_{m}}^{DL}(i) R_{RU_{m}}^{DL}(i) - q_{BR}^{DL}(i) R_{BR}^{DL}(i)  \Big] \notag \\
+ \sum_{i=1}^{N} \alpha^{DL} (i) \Big[ \sum_{l=1}^M q_{RU_{m}}^{DL}(i) + \sum_{l=1}^M q_{BU_{m}}^{DL}(i) + q_{BR}(i) -1 \Big] \notag \\
+  \sum_{i=1}^{N} \sum_{XY} \eta_{XY}^{DL}(i) \Big[ q_{XY}^{DL}(i)-1 \Big] - \sum_{i=1}^{N} \sum_{XY} \xi_{XY}^{DL}(i) q_{XY}^{DL}(i)  \notag
\end{IEEEeqnarray}
where $\lambda_{2} , \alpha^{DL} (i), \eta_{XY}^{DL}(i), \xi_{XY}^{DL}(i)$ are Lagrange multipliers.

We take the UL optimization as an example, the KKT conditions include the following:

(1) Stationary condition: 
\begin{IEEEeqnarray}{C}
\frac{\partial {\cal L}^{UL}}{\partial q_{U_{m}B}^{UL}(i)} =0, \forall i,m; \frac{\partial {\cal L}^{UL}}{\partial q_{U_{m}R}^{UL}(i)} =0, \forall i,m; \frac{\partial {\cal L}^{UL}}{\partial q_{RB}^{UL}(i)} =0 \notag
\end{IEEEeqnarray}

(2) Primal feasibility condition: constraints A2 and A3

(3) Dual feasibility condition:
\begin{IEEEeqnarray}{C}
\eta_{YX}^{UL}(i) \geq 0, \xi_{YX}^{UL}(i) \geq 0, YX \in {\cal S}^{UL}, \forall i
\end{IEEEeqnarray}

(4) Complementary slackness:
\begin{IEEEeqnarray}{l}
\eta_{YX}^{UL}(i) \Big[ q_{YX}^{UL}(i)-1 \Big]=0, YX \in {\cal S}^{UL}, \forall i \\
\xi_{YX}^{UL}(i) q_{YX}^{UL}(i)=0, YX \in {\cal S}^{UL}, \forall i 
\end{IEEEeqnarray}

From stationary condition, we can get 
\begin{IEEEeqnarray}{l}
- \frac{1}{N}  R_{U_{m}B}^{UL}(i) + \alpha^{UL} (i) +  \eta_{U_{m}B}^{UL}(i) - \xi_{U_{m}B}^{UL}(i) =0 , \forall i,m \label{app1}\\
\lambda_{1} \frac{1}{N} R_{U_{m}R}^{UL} (i) + \alpha^{UL} (i) +  \eta_{U_{m}R}^{UL}(i) - \xi_{U_{m}R}^{UL}(i) =0 , \forall i,m \label{app2}\\
- (1+\lambda_{1}) \frac{1}{N}  R_{RB}^{UL}(i) + \alpha^{UL} (i) +  \eta_{RB}^{UL}(i) - \xi_{RB}^{UL}(i) =0 , \forall i \label{app3}
\end{IEEEeqnarray}

Without loss of generality, if $q_{U_{m}B}^{UL*}(i)=1$ , then $q_{U_{j}B}^{UL*}(i)=0,\forall j \neq m$, $ q_{U_{j}R}^{UL*}(i)=0, \forall j$ and $p_{RB}^{UL*}(i)=0$. From complementary slackness, we obtain 
\begin{IEEEeqnarray}{ll}
\xi_{U_{m}B}^{UL}(i) = 0 ; & \eta_{U_{j}B}^{UL}(i) = 0, \forall j \neq m \notag \\
\eta_{U_{j}R}^{UL}(i) =0, \forall j ; & \eta_{RB}^{UL}(i)=0 \notag
\end{IEEEeqnarray}
Thus from \eqref{app1} \eqref{app2} \eqref{app3} and dual feasibility condition, we get
\begin{IEEEeqnarray}{l}
R_{U_{m}B}^{UL}(i)-R_{U_{j}B}^{UL}(i) \geq 0, \forall j \neq m \\
R_{U_{m}B}^{UL}(i)+\lambda_{1} R_{U_{j}R}^{UL}(i) \geq 0, \forall j  \\
R_{U_{m}B}^{UL}(i)-(1+\lambda_{1})  R_{RB}^{UL}(i)  \geq 0
\end{IEEEeqnarray}

Follow the similar process, if $q_{U_{m}R}^{UL*}(i)=1$, 
\begin{IEEEeqnarray}{l}
-\lambda_{1} ( R_{U_{m}R}^{UL}- R_{U_{j}R}^{UL} ) \geq 0, \forall j \neq m \\
-\lambda_{1} R_{U_{m}R}^{UL} - R_{U_{j}B}^{UL}(i) \geq 0, \forall j  \label{contra1}\\
-\lambda_{1} R_{U_{m}R}^{UL}-(1+\lambda_{1})  R_{RB}^{UL}(i)  \geq 0
\end{IEEEeqnarray}

If $q_{RB}^{UL*}(i)=1$,
\begin{IEEEeqnarray}{l}
(1+\lambda_{1})  R_{RB}^{UL}(i) - R_{U_{j}B}^{UL}(i)  \geq 0, \forall j \label{contra2} \\
(1+\lambda_{1})  R_{RB}^{UL}(i) + \lambda_{1} R_{U_{j}R}^{UL}\geq 0, \forall j  
\end{IEEEeqnarray}

Thus we get the necessary condition for the optimal selection in the $i$-th slot, which leads to the Proposition 1. In particular, when $-1 < \lambda_{1} < 0$, the criterion is UL case I. While if $\lambda_{1} \geq 0$, it leads to contradiction with \eqref{contra1}, such that there will be no input of the buffer in UL, while output is not necessary. If $\lambda_{1} \leq -1$, it will lead to contradiction with \eqref{contra2}, such that no output will be selected and no input should happen, otherwise bits will be trapped in the buffer. In summary, $\lambda_{1} \geq 0$ or $\lambda_{1} \leq -1$ leads to UL case II.

\section{}\label{appb}
The proof skeleton for Proposition 2 is similar to that of Propostion 1 in Appendix \ref{appa}. In this case, we consider the Lagrangian function of the relaxed problem with KKT condition for NODBA protocol.
\begin{IEEEeqnarray}{l}
{\cal L}= - \frac{1}{N} \sum_{i=1}^{N} \Big[  
 \sum_{m} \sum_{l \neq m} q_{(m,l)}^{T1}(i) R_{BU_{l}}^{T1}(i) + q_{RB}^{T3}(i) R_{RB}^{T3}(i) \notag \\
+ \sum_{m} \sum_{l \neq m} q_{(m,l)}^{T2}(i) \big[ R_{U_{m}B}^{T2}(i) + R_{RU_{l}}^{T2}(i)   \big] \Big] \notag \\
+\lambda_{3} \frac{1}{N} \sum_{i=1}^{N}  \Big[ \sum_{m} \sum_{l \neq m} q_{(m,l)}^{T1}(i) R_{U_{m}R}^{T1}(i) - q_{RB}^{T3}(i) R_{RB}^{T3}(i) \Big] \notag \\
+\lambda_{4} \frac{1}{N} \sum_{i=1}^{N} \Big[ q_{BR}^{T4}(i) R_{BR}^{T4}(i) - \sum_{m} \sum_{l \neq m} q_{(m,l)}^{T2}(i) R_{RU_{l}}^{T2}(i) \Big] \notag \\
+ \sum_{i=1}^{N} \alpha (i) \Big[ \sum_{m} \sum_{l \neq m} q_{(m,l)}^{T1}(i) + \sum_{m} \sum_{l \neq m} q_{(m,l)}^{T2}(i) 
+ q_{RB}^{T3}(i) +q_{BR}^{T4}(i) - 1 \Big] \notag \\
+ \sum_{i=1}^{N} \sum_{m} \sum_{l \neq m} \Big\{ \eta_{(m,l)}^{T1}(i) \Big[ q_{(m,l)}^{T1}(i)-1 \Big] - \xi_{(m,l)}^{T1}(i) q_{(m,l)}^{T1}(i) \Big\} \notag \\
+ \sum_{i=1}^{N} \sum_{m} \sum_{l \neq m} \Big\{ \eta_{(m,l)}^{T2}(i) \Big[ q_{(m,l)}^{T2}(i)-1 \Big] - \xi_{(m,l)}^{T2}(i) q_{(m,l)}^{T2}(i) \Big\} \notag \\
+ \sum_{i=1}^{N} \Big\{ \eta_{RB}^{T3}(i) \Big[ q_{RB}^{T3}(i)-1 \Big] -\xi_{RB}^{T3}(i) q_{RB}^{T3}(i) \Big\} 
+ \sum_{i=1}^{N} \Big\{ \eta_{BR}^{T4}(i) \Big[ q_{BR}^{T4}(i)-1 \Big] -\xi_{BR}^{T4}(i) q_{BR}^{T4}(i) \Big\} \notag
\end{IEEEeqnarray}
where $\lambda_{3}, \lambda_{4} , \alpha (i), \eta_{(m,l)}^{T1}(i), \xi_{(m,l)}^{T1}(i), \eta_{(m,l)}^{T2}(i), \xi_{(m,l)}^{T2}(i)$, $\eta_{RB}^{T3}(i), \xi_{RB}^{T3}(i), \eta_{BR}^{T4}(i)$ and $\xi_{BR}^{T4}(i)$ are Lagrange multipliers.

\bibliographystyle{IEEEtranTCOM}
\bibliography{IEEEabrv,Liurkbibfile}

\end{document}